\documentclass{aa}
\usepackage{times}
\usepackage{graphics}
\usepackage{xspace}
\usepackage{epsfig}
\usepackage{jwaabib}
\usepackage{rotating}
\usepackage{dcolumn}

\newcommand{\myrule}{\rule[-0.1cm]{0.cm}{0.5cm}}

\newcommand{\yygem}{YY\,Gem\xspace}

\begin{document}


\title{Castor\,A and Castor\,B resolved \\ in a simultaneous {\em Chandra} and {\em XMM-Newton} observation}

\author{B. Stelzer\inst {1} \and V. Burwitz\inst {2}}

\institute{$^1$ INAF- Osservatorio Astronomico di Palermo,
  Piazza del Parlamento 1,
  I-90134 Palermo,
  Italy \\ 
  $^2$ Max-Planck-Institut f\"ur extraterrestrische Physik,
  Postfach 1312, 
  D-85741 Garching,
  Germany}

\offprints{B. Stelzer}
\mail{B. Stelzer, stelzer@astropa.unipa.it}
\titlerunning{Castor\,A and Castor\,B resolved in X-rays}

\date{Received $<$24-12-02$>$ / Accepted $<$20-02-03$>$}

\abstract{
We present a simultaneous {\em Chandra} and {\em XMM-Newton} observation of
the Castor sextett, focusing on Castor\,A and Castor\,B, 
two spectroscopic binaries with early-type primaries.
Our study represents the first unambiguous X-ray detection of all three visual
components in the Castor sextett making use of the unprecedented spatial
resolution of {\em Chandra}. Of the present day X-ray instruments only
{\em Chandra} can isolate the X-ray lightcurves and spectra of Castor\,A
and~B (angular separation $\sim 4^{\prime\prime}$). We compare the
{\em Chandra} observation with {\em XMM-Newton} data obtained simultaneously.
Albeit not able to resolve Castor\,A and Castor\,B from each other, 
the higher sensitivity of {\em XMM-Newton} allows for a quantitative analysis
of their combined high-resolution spectrum. He-like
line triplets are used to examine the temperature and the density in the corona
of Castor\,AB. The oxygen triplet provides a density of  
$n_{\rm e} = (0.5...1)\,10^{10}\,{\rm cm^{-3}}$, typical for stellar coronae.
The analysis for the neon triplet results in much higher densities. By means 
of a simulated RGS spectrum we 
estimate the contaminating effect of iron lines to the neon triplet. 
The temporal variability of Castor\,AB is studied using data collected with the
European Photon Imaging Camera onboard {\em XMM-Newton}. Strong
flare activity is observed with typical rise times of $\sim 10$\,min and 
exponential decays which are by a factor of $2-3$ slower than the rise. 
Combining the data acquired
simultaneously with {\em Chandra} and {\em XMM-Newton} each flare can be
assigned to its host. Thus we verify that both Castor\,A and Castor\,B
exhibit flares.
Our comparison with the conditions of the coronal plasma
of other stars shows that Castor\,AB behave like typical late-type
coronal X-ray emitters supporting the common notion that the late-type
secondaries within each spectroscopic binary are the sites of the X-ray
production.
}

\maketitle

\section{Introduction}\label{sect:intro}

Castor is a sextett system at a distance of $16$\,pc, and 
composed of three visual stars
each of which is by itself a spectroscopic binary. Castor\,C (= YY\,Gem) 
consists of two late-type stars (dM1e + dM1e) and is separated from 
Castor\,AB by $\sim 74^{\prime\prime}$. The primaries in both Castor\,A
and Castor\,B are of spectral type A (A1\,V and A5\,V, respectively). 
Neither strong stellar winds
(typical for early-type stars) nor magnetic activity (typical for
late-type stars) are thought to produce X-ray emission in early A-type
stars. Nevertheless, observations by {\em EXOSAT} allowed to separate 
two sources of X-ray emission, \yygem and the visual binary of 
Castor AB (\cite{Pallavicini90.1}). 
Castor AB was also marginally resolved from \yygem by {\em ASCA}
(see \cite{Gotthelf94.1}).
The most likely explanation is that 
the presumably late-type spectroscopic companions of Castor\,A and/or 
Castor\,B are responsible for the X-ray emission.
The companion of Castor\,A is most likely a late-K star in a 9.21\,d
eccentric orbit, while Castor\,B's companion seems to be an early-M dwarf
with a 2.93\,d circular orbit 
(\cite{Guedel96.1}, \cite{Heintz88.1}).
\citey{Schmitt94.1} have argued in favor of Castor\,A as the X-ray source
because they detected microwaves from Castor\,A but not from Castor\,B. 
The latter was, however, subsequently detected as a radio source as well
(\cite{Guedel96.1}).
X-ray and microwave emission are expected to be associated with each other
because both are generated as a result of magnetic energy release in flares.
In contrast to the Sun, stars are known to be microwave emitters also
during phases of quiescence. X-ray and microwave emission in stars have
been found to be strongly correlated (\cite{Benz94.1}, \cite{Guedel93.1}). 

In a recent {\em XMM-Newton} observation \citey{Guedel01.1} found that
all three visual binaries in the sextett are X-ray sources. 
The angular separation of Castor\,A and\,B is $\sim 4^{\prime\prime}$, at the
limit of the spatial resolving power of {\em XMM-Newton}. 
The superior resolution of {\em Chandra} now allows us  
to present an unambiguous X-ray detection of both Castor\,A and Castor\,B, and 
for the first time to isolate the X-ray spectrum and lightcurves of both 
binaries. 
A contemporaneous {\em XMM-Newton} observation provides higher signal-to-noise 
at both medium- and high-spectral resolution. 
To date high-resolution X-ray spectra have been presented for about a dozen
late-type stars (e.g. \cite{Guedel01.1}, \cite{Ness01.1}, 
\cite{Mewe01.1}, \cite{Raassen02.1}, \cite{Stelzer02.1}, \cite{Ness02.1}),
providing for the first time direct access to the physical conditions in 
stellar coronae by measuring fluxes of individual emission lines. 
The relative strength of lines from different elements shed light on the
temperature, density and abundances in the emitting region. 
Enlarging the data base of high-resolution studies of stars is important
to distinguish which of the observed properties are typical for stellar coronae
and which of the objects are peculiar. Observing a variety of stars
with different spectral type, multiplicity, rotation, and age will eventually
allow to disentangle the influences of these parameters on stellar activity. 
The combination of our simultaneous {\em XMM-Newton} and {\em Chandra} 
observation provides a unique data set with all the capabilities 
of the present-day generation of X-ray instruments, and enables 
a detailed study of the characteristics of three coronal X-ray sources in 
the Castor system.

In Sect.~\ref{sect:obs} we introduce the observations. The data reduction 
is described in Sect.~\ref{sect:data}. 
We present in brief our analysis of the medium-resolution spectrum 
(Sect.~\ref{sect:medres_spec}). 
Modelling the high-resolution spectra obtained with both {\em Chandra} and {\em XMM-Newton}, 
and interpretation of the results 
is the major purpose of this paper (Sect.~\ref{sect:highres_spec}). 
Therefore, Sect.~\ref{sect:medres_spec} 
does not aim at an exhaustive discussion of the medium-resolution spectrum of Castor\,AB. 
Rather it is introduced because  
of the additional information it provides for a better understanding
of the high-resolution spectrum which is used along this paper. 
A study of the X-ray variability of Castor\,A and~B based on the X-ray lightcurves 
is given in Sect.~\ref{sect:lcs}.
In Sect.~\ref{sect:act} we compare the
activity of Castor\,A and~B to other active stars, 
and in Sect.~\ref{sect:summary} we summarize our results.

\section{Observations}\label{sect:obs}

We have obtained deep X-ray observations of the Castor system with
both {\em Chandra} and {\em XMM-Newton} (Obs-IDs 28 and 0112880801,
respectively). The observations took place
simultaneously on Sep 29/30, 2000. {\em Chandra} was used in the
the LETGS configuration, i.e. the Low Energy Transmission Grating (LETG)
was combined with the High Resolution Camera for Spectroscopy (HRC-S).
{\em XMM-Newton} allows to perform high-resolution spectroscopy with the
Reflection Grating Spectrometer (RGS) and at the same time CCD imaging
and spectroscopy with the European Photon Imaging Camera (EPIC). 
During the observation EPIC was operated in full-window mode with the thick 
filter inserted due to the optical brightness of Castor ($V \sim 1.6$\,mag).
For the same reason the Optical Monitor was in closed position.

The prime target of this campaign
was YY\,Gem (= Castor\,C), discussed by \citey{Stelzer02.1} where we also 
presented the observing log. The instrumental setup
allows to analyse also the temporal and spectral behavior of the other
two visual components of the sextett, Castor\,A and Castor\,B. In this paper we
focus on the X-ray emission of these two binary stars.

\section{Data Reduction}\label{sect:data}

\begin{figure}
\begin{center}
\resizebox{8cm}{!}{\includegraphics{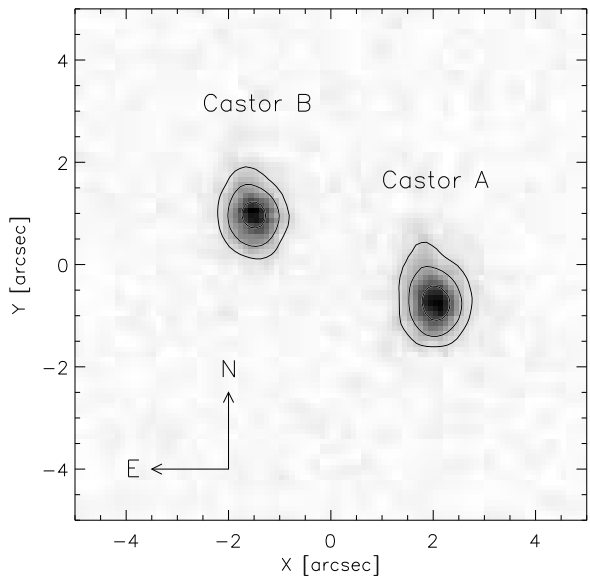}}
\caption{{\em Chandra} zeroth order LETGS image of Castor\,A and Castor\,B. Their separation in this image is $3.8^{\prime\prime}$.}
\label{fig:letgs_image}
\end{center}
\end{figure}

\begin{figure*}
\begin{center}
\resizebox{16cm}{!}{\includegraphics{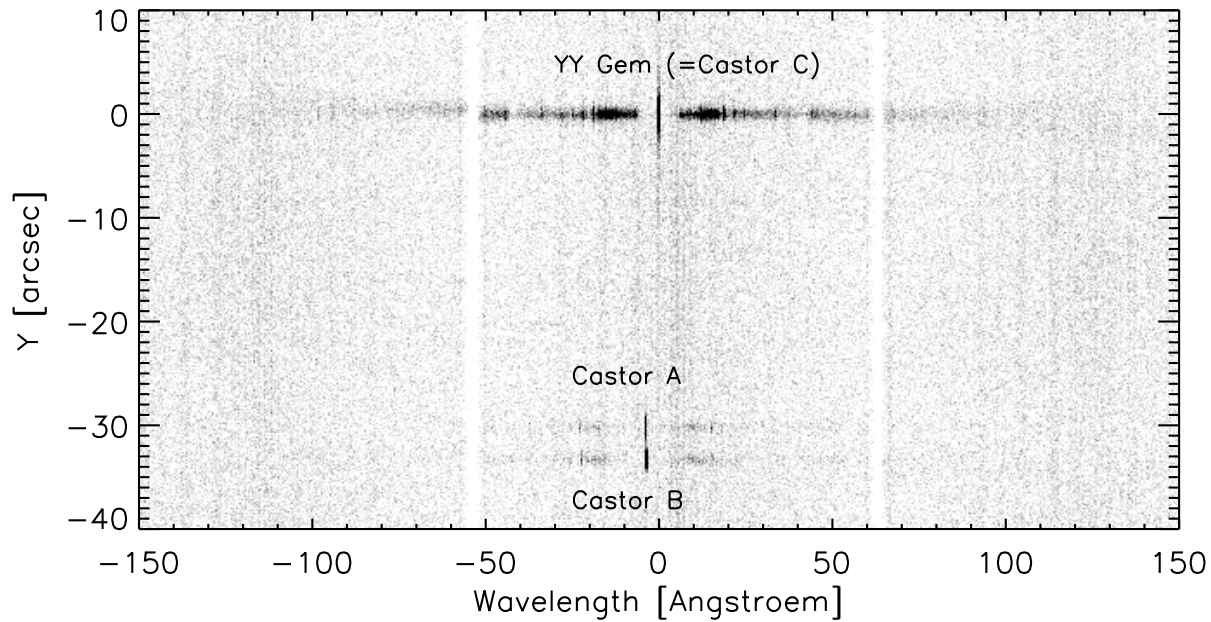}}
\caption{Raw LETGS spectrum of the Castor system.}
\label{fig:raw_spec}
\end{center}
\end{figure*}

Fig.~\ref{fig:letgs_image} shows the zeroth order LETGS image of Castor\,A and~B.
Their separation on the LETGS image is
$3.8^{\prime\prime}$ and the position angle is $64.9^\circ$, 
in agreement with the value expected from the optical ephemeris 
(\cite{Heintz88.1}).
With the LETGS both the dispersed spectrum of Castor\,A and Castor\,B and their
lightcurves are separable. 
In Fig.~\ref{fig:raw_spec} the raw LETGS spectrum of the Castor system is
displayed. 
To extract the LETGS spectrum for Castor\,A and Castor\,B we followed the 
CIAO 2.2 science threads\footnote{http://asc.harvard.edu/ciao/threads/spectra-multi-hrc/} for grating spectroscopy with multiple sources. 
Three sources are detected with {\it tgdetect},
corresponding to YY\,Gem, Castor\,A, and Castor\,B. 
We selected a rectangular box of $0.0005^\circ$ width 
in the cross-dispersion
direction defining the source spectrum of each Castor\,A and Castor\,B. 
The background was extracted from two larger 
rectangular regions above and below the source spectrum, and excluding 
respectively the other two X-ray sources.
Lightcurves for Castor\,A and Castor\,B are extracted from a 
circle of radius $2^{\prime\prime}$ at their zeroth order position. The 
background is estimated from an average of the counts in four 
square-shaped regions arranged symmetrically around the source. After
background subtraction we find a total of $1989 \pm 46$\,cts and 
$1667 \pm 43$\,cts for Castor\,A and~B respectively. The exposure time
of {\em Chandra} was $57.9$\,ksec.  

With {\em XMM-Newton} Castor\,A and~B are not resolved, but its instruments
provide higher sensitivity allowing for a detailed and quantitative analysis. 
The RGS data were reduced with the standard pipeline {\it rgsproc}
of the {\em XMM-Newton} Science Analysis System (SAS), version 5.3. 
With help of the {\it bkgexclude} parameter we 
made sure that YY\,Gem is excluded from the background extraction region.
We extracted the first order total spectrum of Castor\,AB 
(source plus background) and a background spectrum for analysis with the 
CORA \footnote{CORA is provided by J.-U. Ness and available at http://www.hs.uni-hamburg.de/DE/Ins/Per/Ness/Cora/index.html} line fitting program (see Sect.~\ref{sect:highres_spec}). 
The RGS response matrix was generated with 6500 energy bins. 

The data collected by the European Photon Imaging Camera (EPIC) onboard
{\em XMM-Newton} was also analysed with the 
standard {\em XMM}-SAS tools, version 5.3. 
The position of the X-ray source representing Castor\,AB was determined 
by searching in an iterative way for the photon centroid at the expected 
position. Lightcurves and spectra for Castor\,AB were extracted from a 
$25^{\prime\prime}$ 
sized circle centered on their X-ray 
position. This value for the extraction radius corresponds to the point 
where the integrated radial distribution of counts around the position of 
Castor\,AB flattens out. This way we ensure that YY\,Gem does not contribute
significantly to the selected photons. 
A $\sim 1$\,ksec long time interval near the end of the
{\em XMM-Newton} pointing was cut out because of higher background
(see Fig.~\ref{fig:lcs}). The total useful exposure time with EPIC pn
amounts to 48.6\,ksec.

\section{Medium-Resolution Spectrum: EPIC pn}\label{sect:medres_spec}

Since this paper concentrates on the specific new achievements enabled by 
the high spatial and spectral resolution of {\em Chandra} and {\em XMM-Newton} 
the EPIC spectrum will only briefly be introduced. We discuss only the pn 
spectrum, which provides the highest sensitivity among the EPIC instruments. 

For the spectral fitting we select photons from a circular region as described
in Sect.~\ref{sect:data}. A background spectrum is obtained from a nearby
source-free region of the same area. We restrict the analysis to energies
above $0.3$\,keV, because the spectral response of pn is not well understood
for lower energies. 
We use the pn redistribution matrix for single and double event patterns
released in April 2002 together with an ancilliary response file
generated in the course of our data reduction process. 
Modelling of the spectrum is performed with XSPEC version 11.2.0. 

Considerable emission is seen in the pn spectrum up to comparatively high
energies. The statistics are good enough to visually identify an Fe feature 
at $\sim 6.7$\,keV. The shape of the spectrum is best approximated by a 
three-temperature (3-T) model describing thermal
emission from a hot, optically thin plasma (three VMEKAL models). 
In the selected energy range absorption is negligible. 
A model with solar abundances gives $\chi^2_{\rm red}=1.65$ for $528$\,dof,  
and leaves substantial residuals especially near $10$\,\AA (neon) 
and $22$\,\AA (oxygen). An acceptable solution 
($\chi^2_{\rm red} \sim 1$) is found for subsolar abundances of iron, 
oxygen, and neon. All other elements do not have strong enough lines in 
the sensitive spectral range of EPIC to change the fit result significantly. 
We summarize the best fit parameters for the 3-T model with variable
abundances in Table~\ref{tab:epic}.
\begin{table}
\begin{center}
\caption{Best fit parameters of a 3-temperature VMEKAL model 
describing the EPIC pn spectrum of Castor\,AB.}
\label{tab:epic}
\begin{tabular}{rrrrr}\hline\hline \noalign{\smallskip}
\multicolumn{1}{c}{$\chi^2_{\rm red} (dof)$}  & \multicolumn{1}{c}{$kT_1$}      & \multicolumn{1}{c}{$kT_1$} & \multicolumn{1}{c}{$kT_3$} & \multicolumn{1}{c}{$\lg{EM_1}$}       \\ \noalign{\smallskip} \hline \noalign{\smallskip} 
                                               & \multicolumn{1}{c}{[keV]}      & \multicolumn{1}{c}{[keV]}  & \multicolumn{1}{c}{[keV]}   & \multicolumn{1}{c}{${\rm [cm^{-3}]}$} \\ \noalign{\smallskip} \hline \noalign{\smallskip}
$1.15 (525)$                                   & \myrule $0.27^{+0.01}_{-0.01}$ & $ 0.78^{+0.03}_{-0.03}$    & $1.68^{+0.12}_{-0.12}$      & $51.49^{+0.04}_{-0.03}$              \\ \noalign{\smallskip} \hline\hline \noalign{\smallskip} 
\multicolumn{1}{c}{$\lg{EM_2}$}       & \multicolumn{1}{c}{$\lg{EM_3}$}       & \multicolumn{1}{c}{$O$}               & \multicolumn{1}{c}{$Ne$} & \multicolumn{1}{c}{$Fe$} \\ \noalign{\smallskip} \hline \noalign{\smallskip} 
\multicolumn{1}{c}{${\rm [cm^{-3}]}$} & \multicolumn{1}{c}{${\rm [cm^{-3}]}$} &                         &                          &                          \\ \noalign{\smallskip} \hline \noalign{\smallskip}
$51.33^{+0.05}_{-0.05}$               & $51.23^{+0.05}_{-0.06}$               & \myrule $0.59^{+0.05}_{-0.06}$        & $0.64^{+0.22}_{-0.28}$   & $0.61^{+0.06}_{-0.07}$   \\ \noalign{\smallskip} \hline\hline \noalign{\smallskip} 
\end{tabular}
\end{center}
\end{table}

\section{High-Resolution Spectra: LETGS and RGS}\label{sect:highres_spec}

\begin{figure*}
\begin{center}
\resizebox{18cm}{!}{\includegraphics{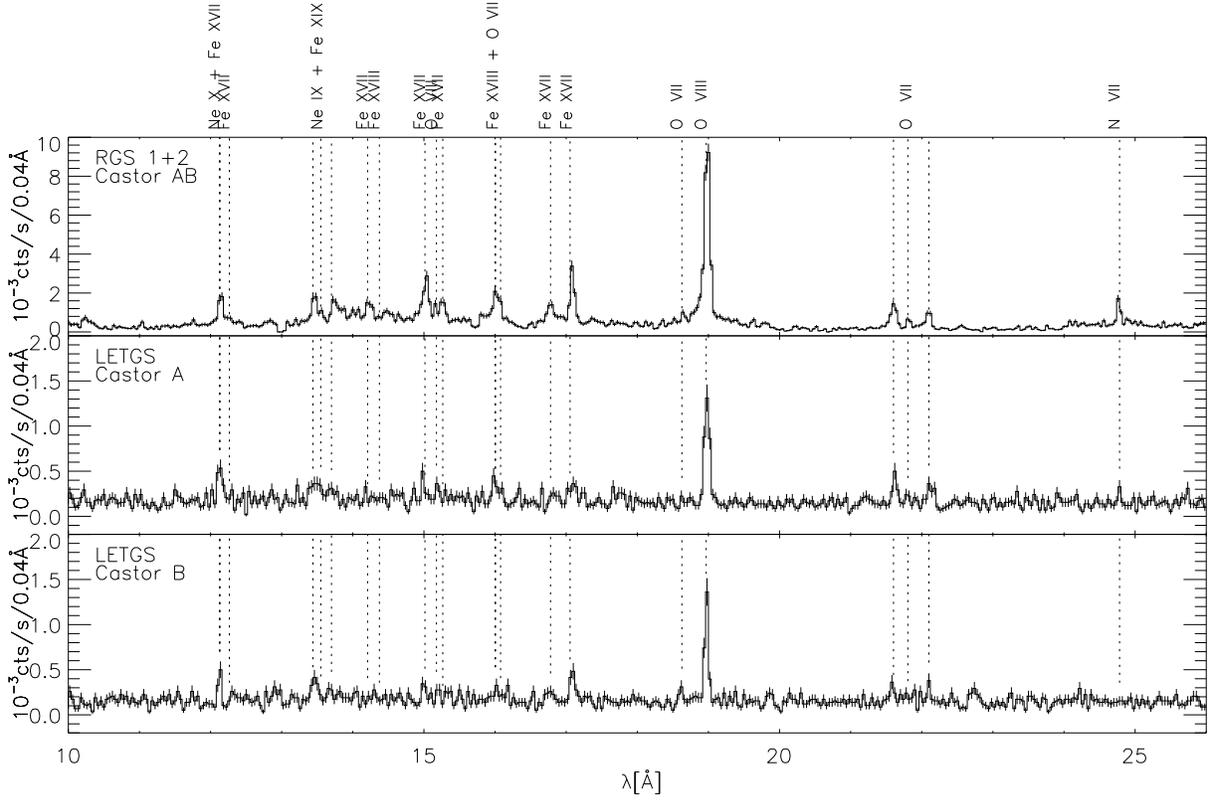}}
\caption{First order RGS count rate spectrum (top) of Castor\,AB and LETGS count rate spectra of Castor\,A and~B (middle and bottom). The counts from RGS\,1 and RGS\,2 have been added. Note, that due to the failed CCDs on each RGS the relative strengths of some lines appear distorted in the combined spectrum. For the LETGS left and right side have been added. Outside the displayed range only one emission line is detected, the Ly$\alpha$ line of C\,VI at $33.7$\,\AA.}
\label{fig:extract_spec}
\end{center}
\end{figure*}

The high-resolution spectrum of both Castor\,A and Castor\,B 
(displayed in Fig.~\ref{fig:extract_spec}) is that of a typical active 
late-type dwarf star. 
The LETGS spectrum, though being quite weak, shows that 
the major emission lines in both stars are from O\,VIII, O\,VII, Ne\,IX,
and Fe\,XVII. 
A quantitative analysis of the LETGS spectrum by measuring fluxes for 
individual lines is subject to large uncertainties because the effective area
of the LETGS for off-axis sources is not well-understood. 
We estimate these uncertainties (a sum of HRC-S and grating efficiency, 
transmission in the spatially inhomogeneous filter) to amount to $\sim 10$\,\%.
This is comparable to or smaller than the statistical errors due to the 
faintness
of both Castor binaries. RGS provides higher sensitivity, but no information
on the location of the X-ray emission within the Castor\,AB system. In view of
the similarity of the LETGS spectrum of Castor\,A and Castor\,B, 
we feel justified to model the combined RGS spectrum, 
and consider it as representative for both stars. 

%
%

In the following we concentrate on the lines from hydrogen- and helium-like ions. 
The triplets of helium-like ions are composed of resonance $r$, 
intercombination $i$, and forbidden $f$ line, and represent 
the most interesting lines in terms of plasma diagnostics:  
Their line ratios $G = (f+i)/r$ and $R = f/i$ 
are sensitive to the electron 
temperature $T_{\rm e}$ and the electron density $n_{\rm e}$ of the emitting 
plasma, respectively (\cite{Gabriel69.1}).
The ratio of hydrogen-like Ly$\alpha$ and helium-like $r$ line provides a measure 
for the coronal temperature,
as it reflects the presence of different ionization stages of the same element. 
In the case of Castor\,AB the only two elements with considerable emission 
above the background
in the triplets and Ly$\alpha$ lines are oxygen and neon. 
The data analysis and results for these two elements are discussed below. 

We measured the line strengths with the CORA line fitting application
(\cite{Ness01.1}). For the LETGS data we used 
Gaussians to represent the individual emission lines. 
For RGS line shapes Lorentzians are a better representation than Gaussians.
All lines are added to a constant describing the local background.

\subsection{Oxygen}\label{subsect:oxygen}

\begin{figure}
\begin{center}
\resizebox{9cm}{!}{\includegraphics{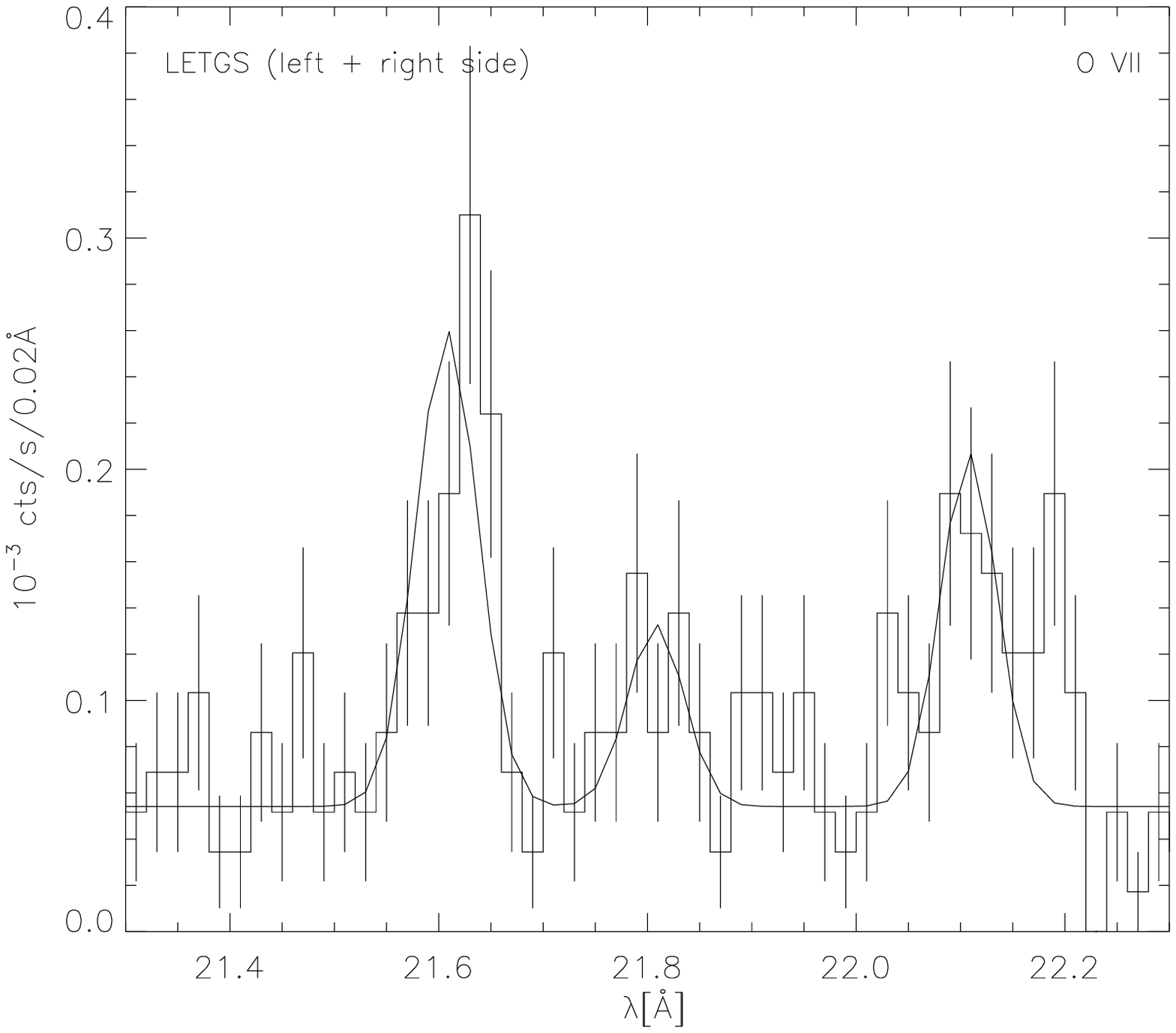}}
\resizebox{9cm}{!}{\includegraphics{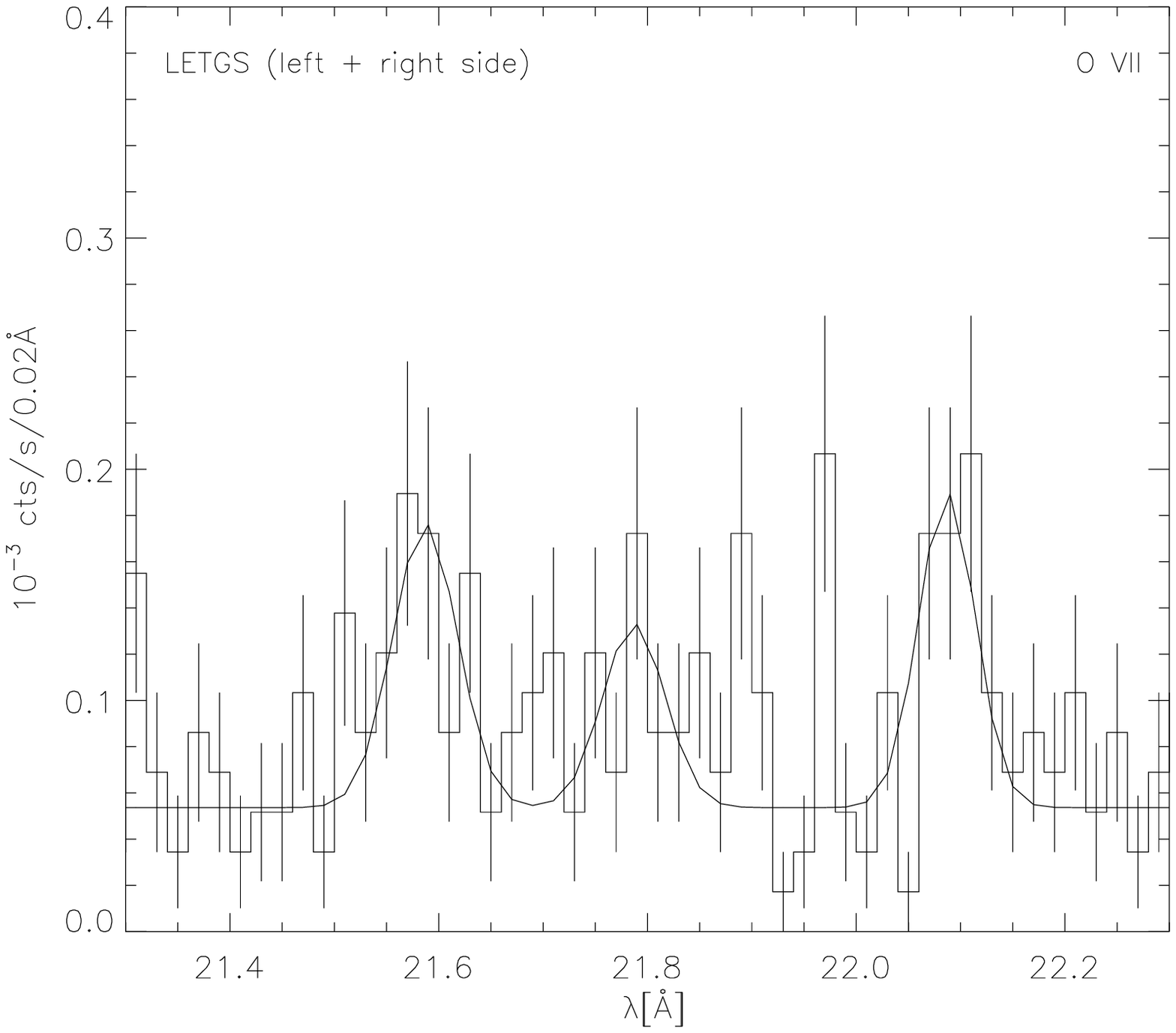}}
\resizebox{9cm}{!}{\includegraphics{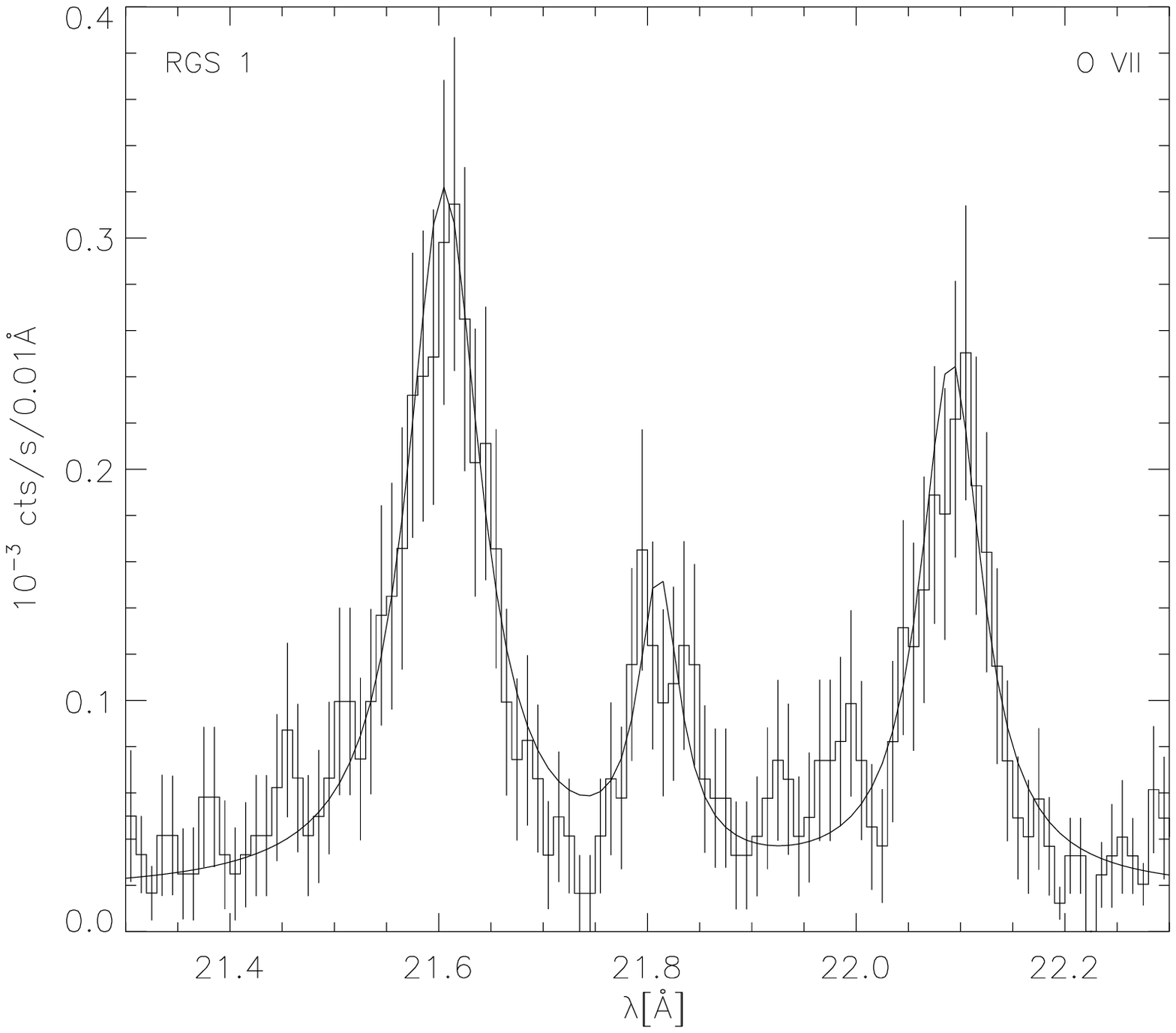}}
\caption{Triplet of helium-like oxygen. From top to bottom: Castor\,A with LETGS, Castor\,B with LETGS, Castor\,AB with RGS. Note the different bin size.}
\label{fig:ovii_triplet}
\end{center}
\end{figure}

In Fig.~\ref{fig:ovii_triplet} we show the LETGS and the RGS spectrum 
in the region of the O\,VII triplet. 
%
%

To increase the signal both sides of the LETGS spectrum were co-added. 
To guide the fit we held the distance between the line centers fixed on the
expected value, and restricted the line widths to $0.03$\,\AA~ corresponding to
the instrumental width of the LETGS. 
This latter restriction assures that random statistical
fluctuations do not influence the line profile. 
Given the large uncertainties related to poor statistics and
low effective area at the off-axis position a conversion of the LETGS 
counts to line fluxes would not warrant meaningful results, and we renounce
on performing this exercise. 

For the quantitative analysis only the RGS is used.
As the CCD in the relevant part of the spectrum has failed on RGS\,2,
only RGS\,1 is available for the analysis of the O\,VII triplet. 
Due to the higher quality of
the RGS spectrum as compared to the LETGS it was possible to leave the line 
centers and line widths free in the fitting process. 
In Table~\ref{tab:lines} we give the line parameters 
(central wavelength, width, number of counts, and photon flux) 
for the O\,VII triplet and O\,VIII Ly$\alpha$ line. 
The flux has been determined from the effective area of the RGS at the 
respective wavelength after modeling of the count spectrum with CORA.  
%
%
\begin{table}
\begin{center}
\caption{Line parameters for the Ly$\alpha$ line of the H-like ion and triplet lines of the He-like ion of oxygen. The best fit result from both RGS and LETGS data are listed. Fluxes were computed only for RGS, because of the uncertain effective area of the LETGS and low signal-to-noise in the spectrum.}
\label{tab:lines}
\begin{tabular}{l|rrrr}\hline
Identif. \myrule & \multicolumn{4}{c}{Line Parameters}     \\
                       & \multicolumn{1}{c}{$\lambda$} & \multicolumn{1}{c}{$\sigma$} & \multicolumn{1}{c}{$I$} & Ph.flux [$10^{-5}$    \\
                       & \multicolumn{1}{c}{[\AA]} & \multicolumn{1}{c}{[\AA]} & \multicolumn{1}{c}{[cts]} & ${\rm cts/s/cm^2}$] \\
\hline
\multicolumn{5}{c}{\bf Castor\,AB -- RGS\,1} \\ \hline
O\,VIII Ly$\alpha$ & $ 18.973$ & $0.087$ & $717.5 \pm 29.4$ & $25.2 \pm 1.0$ \\
O\,VII $r$         & $ 21.605$ & $0.085$ & $218.0 \pm 17.7$ & $ 8.8 \pm 0.7$ \\
O\,VII $i$         & $ 21.811$ & $0.048$ & $ 50.0 \pm 10.0$ & $ 2.1 \pm 0.4$ \\
O\,VII $f$         & $ 22.091$ & $0.071$ & $138.0 \pm 14.3$ & $ 5.7 \pm 0.6$ \\
\hline
\multicolumn{5}{c}{\bf Castor\,A -- LETGS} \\ \hline
O\,VIII Ly$\alpha$ & $18.971$  & $0.033$ & $170.2 \pm 14.1$ & $$ \\
O\,VII $r$         & $21.608$  & $0.030$ & $44.2 \pm 8.1$ & $$ \\ 
O\,VII $i$         & $=21.808$ & $0.027$ & $15.5 \pm 5.8$ & $$ \\ 
O\,VII $f$         & $=22.108$ & $0.027$ & $30.1 \pm 7.2$ & $$ \\ 
\hline
\multicolumn{5}{c}{\bf Castor\,B --  LETGS} \\ \hline
O\,VIII Ly$\alpha$ & $18.963$  & $0.025$ & $127.3 \pm 12.2$ & $$ \\
O\,VII $r$         & $21.587$  & $0.031$ & $27.8 \pm 7.1$ & $$  \\ 
O\,VII $i$         & $=21.787$ & $0.030$ & $17.3 \pm 6.2$ & $$  \\
O\,VII $f$         & $=22.087$ & $0.027$ & $26.7 \pm 6.7$ & $$  \\
\hline
\end{tabular}
\end{center}
\end{table}

\subsection{Neon}\label{subsect:neon}

\begin{figure}
\begin{center}
\resizebox{9cm}{!}{\includegraphics{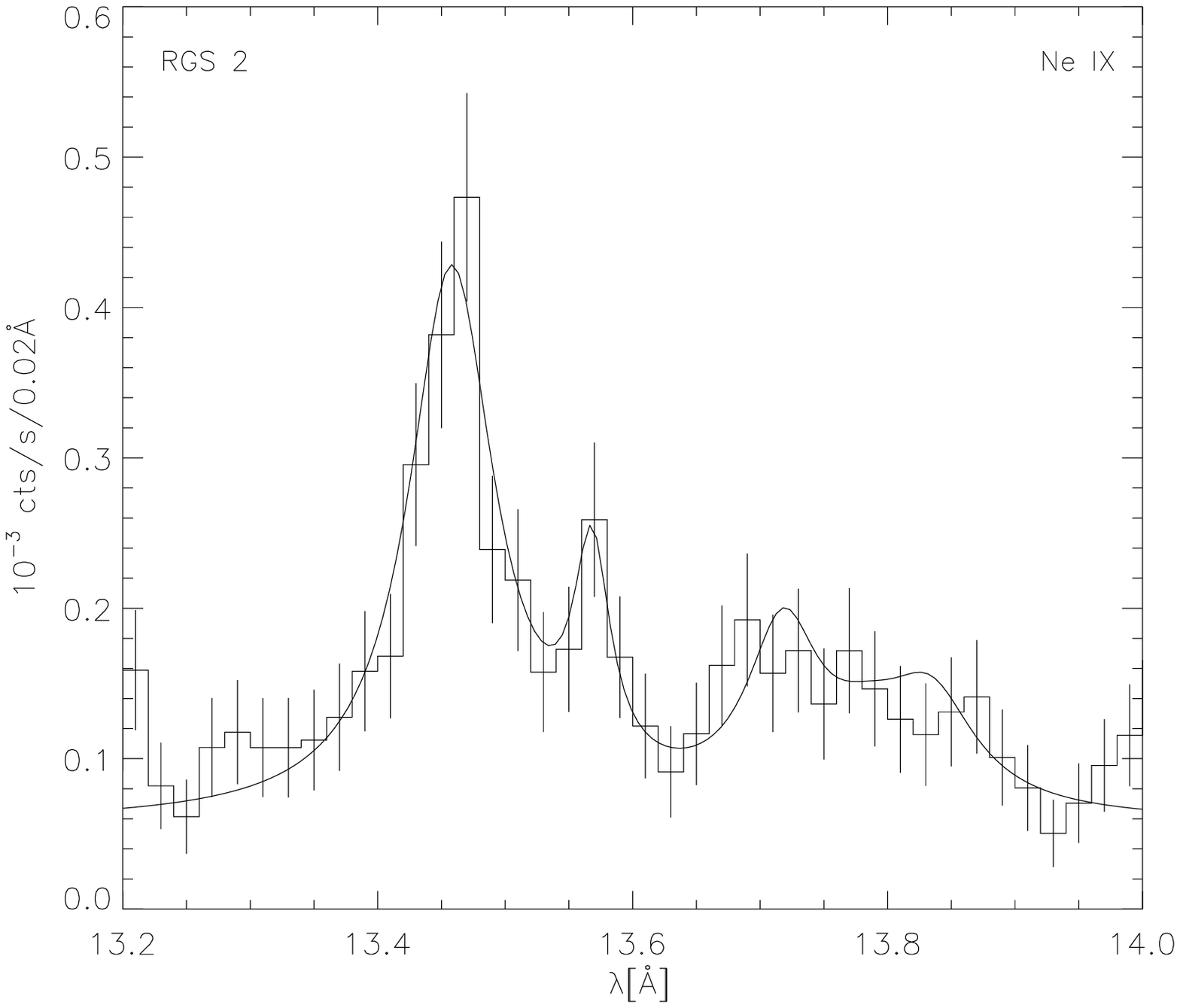}}
\resizebox{9cm}{!}{\includegraphics{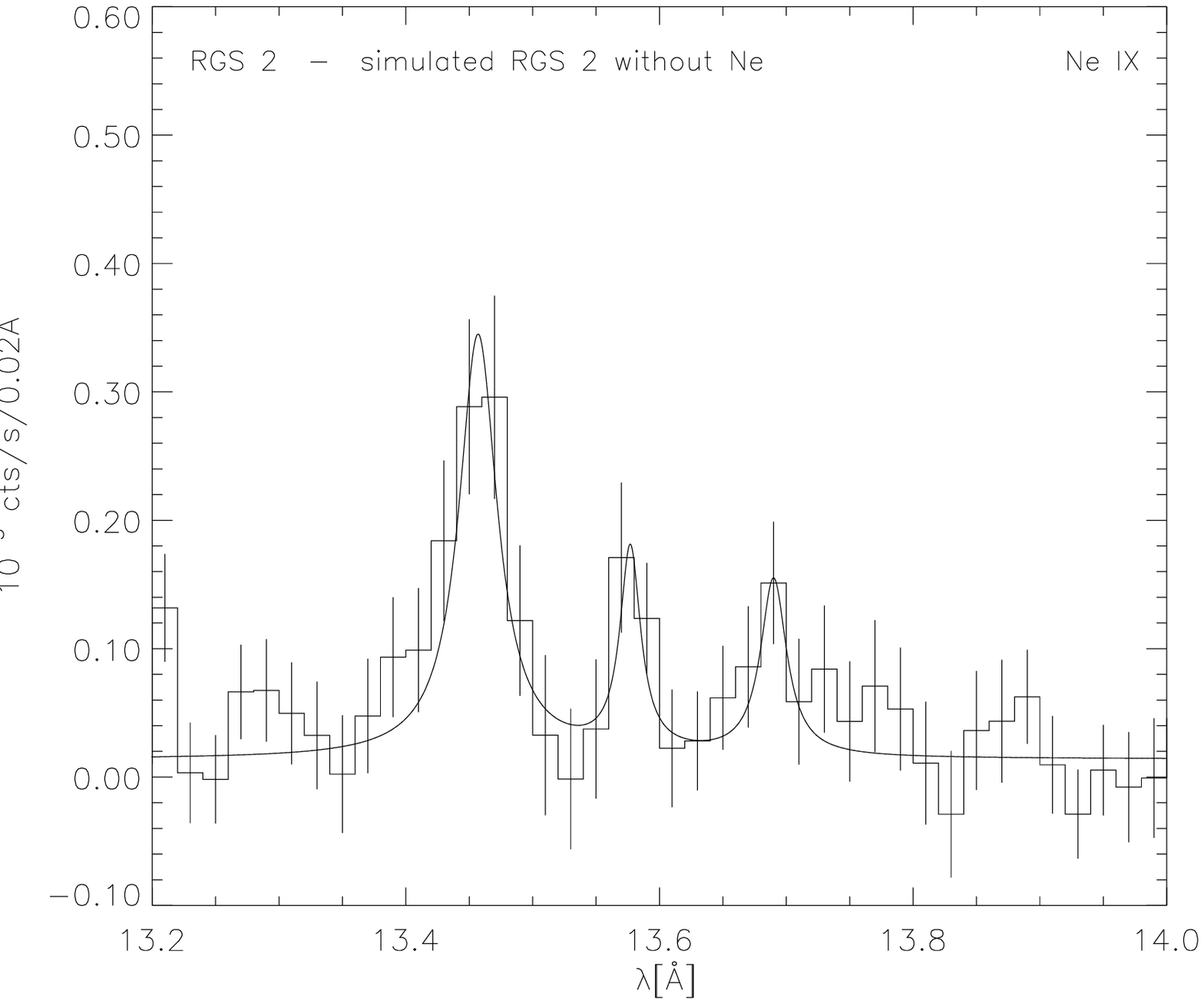}}
\caption{Triplet of helium-like neon. {\em top} - Castor\,AB observed with RGS, fit includes two iron lines that are blended with the triplet (see text in Sect.~\ref{subsect:neon}); {\em bottom} -  RGS spectrum of Castor\,AB after subtraction of a simulated neon-free spectrum, i.e. after removal of the contribution of iron.} 
\label{fig:neix_triplet}
\end{center}
\end{figure}

The neon triplet is more difficult to model, as it is known to be blended 
with a number of iron lines. 
We included additional Lorentzians to account for iron emission
at $13.79$\,\AA~(from Fe\,XIX) and $13.83$\,\AA~(from Fe\,XVII). Further lines
of Fe\,XIX may contaminate the $r$ and the $i$ line of Ne\,IX, but are 
indistinguishable from the latter ones in the data. 
As a first approximation we performed the formal analysis of the neon triplet
in the same way as for oxygen, assuming that the contamination by iron is negligible.
The best fit together with the data is shown in Fig.~\ref{fig:neix_triplet} on
the top. Again only one RGS instrument is available, RGS\,2, because of a
CCD failure in the relevant spectral range on RGS\,1.

To investigate the effect of iron contamination in the neon triplet 
we made use of the information obtained from the medium-resolution
EPIC spectrum (discussed in more detail in Sect.~\ref{sect:medres_spec}).
Modeling the EPIC spectrum provides an estimate for temperature and abundances 
in the X-ray emitting region, but the density can not be well 
constrained. The contribution of iron to the line emission in the critical
spectral range ($\lambda \sim 13.4 ..13.8$\,\AA) can be estimated  
by generating an artificial RGS spectrum for a plasma that has the 
same temperature structure as the EPIC spectrum, but no neon. 
We performed this simulation with XSPEC setting the Ne abundance of the
EPIC 3-T best fit model of Table~\ref{tab:epic} to zero. 
Then we subtracted the simulated neon-free spectrum from the data. The result 
is shown in Fig.~\ref{fig:neix_triplet} (bottom), and should give an idea
about the actual strength of the Ne\,IX triplet. 
Clearly the $r$, $i$, and $f$ lines are more pronounced as compared to 
the pure data, implying that the iron contamination of the data is substantial
rendering any physical parameters derived from the spectrum at these 
wavelengths questionable. 

\subsection{Coronal Density}\label{subsect:porquet}

We derive the electron temperature $T_{\rm e}$ and density $n_{\rm e}$ 
for the oxygen and neon triplets using 
the calculations by \citey{Porquet01.1} for a collision dominated plasma. 
As discussed above the line fluxes for the neon triplet derived from the data
are severely affected by contaminating iron lines.
We continue to include Ne\,IX in the analysis merely with the aim 
to investigate the effect of line blending in this spectral range. 

The assumption of a collisional plasma is justified if
$G \sim 1$, while larger values of $G$ prevail in photo-ionized plasmas.
UV radiation from the star can modify the relative intensities of the
emission lines by populating one level at the expense of another one 
(see \cite{Porquet01.1}). In particular, the upper level of the forbidden
transition can be depopulated to the upper level of the intercombination
transition. For typical late-type coronal X-ray
emitters the UV flux is low and can be neglected. However, for Castor\,A and
Castor\,B the A-type primaries in each of the binaries provide 
substantial UV flux at the position of the X-ray emitting secondary. 
The influence of radiative processes can be expressed in the
radiation temperature $T_{\rm rad}$. 
We compute $T_{\rm rad}$ making use of UV fluxes derived from archived
{\em International Ultraviolet Explorer} (IUE) data of Castor\,A and~B.
In Table~\ref{tab:uv} we tabulate the IUE fluxes of each of the two binaries 
at the wavelengths corresponding to the excitation energy between the upper
levels of the $f$ and the $i$ line for oxygen and neon. 
Making use of the distance ($16$\,pc), 
stellar radius (derived from the Stefan-Boltzmann law), and limb darkening 
(from Table~1 of \cite{Diaz-Cordoves95.1}) 
of Castor\,A and~B we convert the UV flux to intensity and derive 
$T_{\rm rad}$ from the Planck curve for a blackbody emitting at the 
corresponding UV wavelength. The effect of the UV emission at the
position of the secondary is proportionate to the 'dilution factor' $W$,
which depends on the stellar radius and the distance to the UV source, i.e.
the binary separation 
(see \cite{Porquet01.1} for the definition of $W$).
\begin{table}
\begin{center}
\caption{Stellar parameters, IUE flux and radiation temperature of Castor\,A and~B at the wavelengths corresponding to the $f \longrightarrow i$ line transition which is critical for the determination of coronal densities (see text in Sect.~\ref{subsect:porquet}).} 
\label{tab:uv}
\begin{tabular}{llrrrr}
\hline
& & \multicolumn{2}{c}{\bf Castor A} & \multicolumn{2}{c}{\bf Castor B} \\
\hline
\multicolumn{2}{l}{effective temperature (1)}      & \multicolumn{2}{c}{$T_{\rm eff}=10286$\,K} & \multicolumn{2}{c}{$T_{\rm eff}=8842$\,K} \\
\multicolumn{2}{l}{gravity (1)}      & \multicolumn{2}{c}{$\lg{g}=4.2$} & \multicolumn{2}{c}{$\lg{g}=4.0$} \\
\multicolumn{2}{l}{radius}                     & \multicolumn{2}{c}{$R_{\rm *}=2.4\,R_\odot$} & \multicolumn{2}{c}{$R_{\rm *}=3.3\,R_\odot$} \\
\multicolumn{2}{l}{limb darkening coeff. (2)} & \multicolumn{2}{c}{$\epsilon=0.5$} & \multicolumn{2}{c}{$\epsilon=0.5$} \\
\multicolumn{2}{l}{binary separation (3)}          & \multicolumn{2}{c}{$a_{\rm Aa}=0.121$\,AU} & \multicolumn{2}{c}{$a_{\rm Bb}=0.052$\,AU}     \\
\multicolumn{2}{l}{dilution factor }            & \multicolumn{2}{c}{$W=0.002$} & \multicolumn{2}{c}{$W=0.02$}     \\
\hline
                   & & O\,VII & Ne\,IX & O\,VII & Ne\,IX \\
\hline
$\lambda_{\rm fi}$ & [\AA] & 1637   & 1270   & 1637   & 1270 \\
$f_{\rm UV}^*$     & [$10^{10}\,{\rm erg/s/cm^2/\AA}$] & 8      & 4.5    & 8.5    & 4    \\
$T_{\rm rad}$      & [K] & 9950   & 8730   & 9280   & 8250 \\  
\hline
\multicolumn{6}{l}{(1) - \cite{CayreldeStrobel80.1}, (2) - \cite{Diaz-Cordoves95.1},} \\
\multicolumn{6}{l}{(3) - M. G\"udel, priv. comm., $^*$ V. Costa, priv. comm.} \\
\end{tabular}
\end{center}
\end{table}

The case of Castor is complicated by the fact that the UV field and geometrical
situation of both components (A and B) differ slightly from each other, but
only a combined measurement of the X-ray line ratios is available. 
Our RGS measurements for the $G$, $R$, and Ly$\alpha/r$ ratios are 
summarized in Table~\ref{tab:ratios}. 
For the $G$ ratio the effect of UV photo-excitation
is negligible and we can infer an average temperature for Castor\,AB. 
We find $\sim 2$\,MK for oxygen and $\sim 8$\,MK for neon. 
Using these temperatures we calculated the $R$ ratio as a function of density 
for the two components separately: 
(A) adopting the radiation temperature and dilution factor of Castor\,A, 
and (B) for the values of $T_{\rm rad}$ and $W$ of Castor\,B. 
While for neon $T_{\rm rad} < 10000$\,K has no effect on the $R$ ratio,
for oxygen there are slight differences between the two calculations:
We find $n_{\rm e} = 10^{10}\,{\rm cm^{-3}}$ for Castor\,A,
and $n_{\rm e} = 5\,10^{9}\,{\rm cm^{-3}}$ for Castor\,B.
Note that this result was derived under the assumption that both stars
exhibit the same line ratios. Slight differences in the line ratios may
lead to an effect on the density of the same order as the range given above. 
We summarize the results for the temperatures and densities 
in Table~\ref{tab:ratios} together with the line ratios.
%
%
\begin{table}
\begin{center}
\caption{Line ratios derived from the unresolved RGS spectrum of Castor\,AB
and the RGS spectrum cleaned from iron-contribution (see text): 
$G=(f+i)/r$, $R=f/i$, and Ly$\alpha/r$. 
Corresponding electron temperature $T_{\rm e}$ and
electron density $n_{\rm e}$ are derived using the calculations for CIE by 
\protect\citey{Porquet01.1}.} 
\label{tab:ratios}
\begin{tabular}{rrrrr}\hline
\multicolumn{1}{c}{$G$} & \multicolumn{1}{c}{$R$} & \multicolumn{1}{c}{Ly$\alpha/r$} & \multicolumn{1}{c}{$T_{\rm e}$} & \multicolumn{1}{c}{$n_{\rm e}$} \\
                        &                         &                                  & \multicolumn{1}{c}{[MK]}        & \multicolumn{1}{c}{[${\rm 10^{10}cm^{-3}}$]} \\
\hline
\multicolumn{5}{c}{\bf Oxygen} \\
\hline
$0.88 \pm 0.11$         & $2.79 \pm 0.63$         & $2.86 \pm 0.26$                  & $\sim 2$    & $(0.5...1)$ \\
\hline
\multicolumn{5}{c}{\bf Neon} \\
\hline
$0.43 \pm 0.09$         & $1.55 \pm 0.61$         & $1.04 \pm 0.13$                  & $\sim 8$    & $100$ \\
\hline
\multicolumn{5}{c}{\bf Neon, data - simulation} \\
\hline
$0.54 \pm 0.16$         & $1.10 \pm 0.56$         &                                   & $\sim 7$    & $200$ \\
\hline
\end{tabular}
\end{center}
\end{table}

The RGS data of the Ne\,IX triplet results in a very high density. 
Such high densities are untypical, though not generally excluded for active 
stars. Above we have demonstrated that in the case of Castor the neon line
ratios are influenced by blending with iron lines. 
Therefore, a more credible estimate for the density is derived from the 
spectrum after subtraction of the simulated iron contribution.
In Table~\ref{tab:ratios} we 
confront the results obtained directly from the data and after taking
account of the simulation.
It turns out that the temperature and density have not changed by much.
However, the flux of $i$ and $f$, and subsequently any physical parameters 
derived from them, are subject to large uncertainties. 
To summarize, it can be concluded that physical parameters derived from 
the Ne\,IX triplet should always be regarded with caution, 
and this part of the spectrum is not suited for an investigation of the 
conditions in the corona with the RGS instrument. Among the current X-ray
instruments only the HETG is able to sufficiently resolve a stellar spectrum
in this region (see \cite{Ness02.2}).

\section{X-ray Lightcurves}\label{sect:lcs}

%
%
\begin{figure*}
\begin{center}
\resizebox{18cm}{!}{\includegraphics{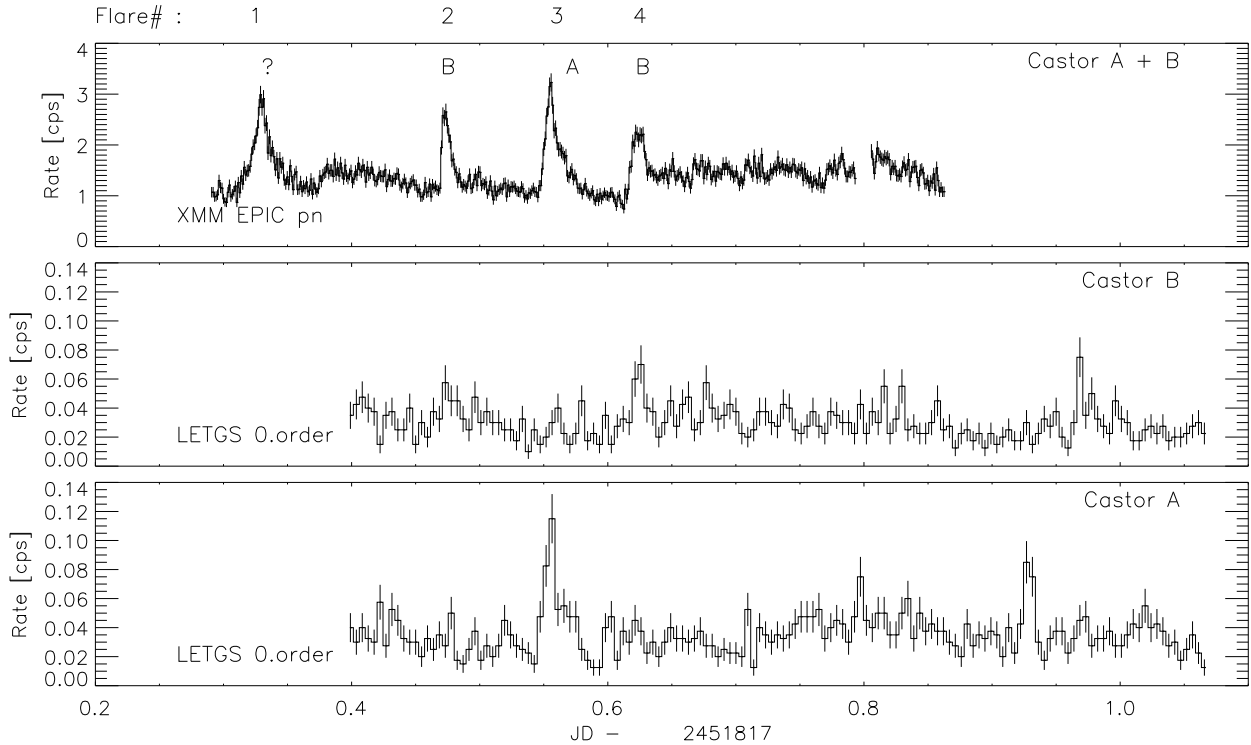}}
\caption{X-ray lightcurves of Castor\,A and~B on Sep 29/30, 2000. {\em top} - {\em XMM-Newton} EPIC pn where Castor\,A and B are unresolved, binsize is 100\,s; {\em middle and bottom} - {\em Chandra} LETGS zeroth order signal of Castor\,B and Castor\,A; binsize is 400\,s.}
\label{fig:lcs}
\end{center}
\end{figure*}

To examine the X-ray variability of Castor\,A and Castor\,B we extracted their 
lightcurves using (A) the {\em Chandra} LETGS zeroth order signal, and (B)
the {\em XMM-Newton} data collected by the EPIC pn.

With EPIC the stars are not resolved, 
so that only a combined lightcurve of Castor\,AB is obtained. 
The EPIC pn lightcurve (Fig.~\ref{fig:lcs}, top panel) 
displays strong variability including four flares within the  
observing time of $\sim 50$\,ksec. The high signal-to-noise of EPIC pn 
allows to estimate characteristic parameters for these flares: 
Rise and decay times, peak luminosity, and energy content are summarized 
in Table~\ref{tab:flareparams}. All flares are of short-duration 
(less than $1$\,h) and of similar strength. While the rise takes place linearly, 
the decay phase follows an exponential resulting from cooling. The decay
times are typically about a factor $2-3$ longer than the rise times. At
peak emission the amplitude has increased by a factor $2-3$. The energy budget
of the flares is listed in columns~5-7 of Table~\ref{tab:flareparams}.
We give the total energy radiated in the flare time interval (as derived from
an integration of the EPIC pn spectrum), the amount of energy above the
quiescent energy during the flares, and the fractional increase of energy
emitted during the flares (see Table caption for a more detailed description).
%
%
\begin{table}
\begin{center}
\caption{Characteristic parameters for flares observed by {\em XMM-Newton} EPIC on Castor\,A and Castor\,B. Rise and decay times are measured from 10\,\% of the amplitude to the peak and vice versa. Energies given in column~5-7 refer to the 0.1-2.4\,keV band. The energy $E_{\rm tot,F}$ was computed by integrating the EPIC pn spectrum in the respective time interval. $E_{\rm F}$ denotes the energy emitted in the flare alone, after subtraction of the average quiescent energy emitted over the duration of the flare. The subsequent column is the fraction of energy radiated due to the flare normalized to the quiescent energy. The flag in the last column identifies the host of the flare: Castor\,A or Castor\,B.}
\label{tab:flareparams}
\begin{tabular}{lrrrrrrl} \\
\hline
\# & $\tau_{\rm rise}$ & $\tau_{\rm decay}$ & $\frac{I_{\rm peak}}{I_{\rm preflare}}$ & $\lg{E_{\rm tot,F}}$ & $\lg{E_{\rm F}}$ & $\frac{E_{\rm F}}{E_{\rm Q}}$ & Host \\ 
   & [min]             & [min]              &                                         & [erg]                & [erg]            &                       &   \\ 
\hline
1  & $18.6$ & $29.8$ & $2.78$ & $32.57$ & $32.12$ & $0.63$ & A$^+$ \\
2  & $ 9.2$ & $15.8$ & $2.42$ & $32.27$ & $31.77$ & $0.55$ & B$^*$ \\
3  & $11.9$ & $36.5$ & $3.11$ & $32.55$ & $32.02$ & $0.51$ & A$^*$ \\
4  & $ 8.6$ & $11.6$ & $2.34$ & $32.47$ & $31.70$ & $0.58$ & B$^*$ \\
\hline
\multicolumn{8}{l}{$^*$ Host assigned from simultaneous, resolved {\em Chandra} lightcurves,} \\
\multicolumn{8}{l}{$^+$ host assigned from photon centroid in MOS image combined with} \\
\multicolumn{8}{l}{information for other flares from {\em Chandra} (see text in Sect.~\ref{sect:lcs}).}
\end{tabular}
\end{center}
\end{table}

\begin{figure}
\begin{center}
\parbox{9cm}{
\parbox{4.5cm}{\resizebox{4.5cm}{!}{\includegraphics{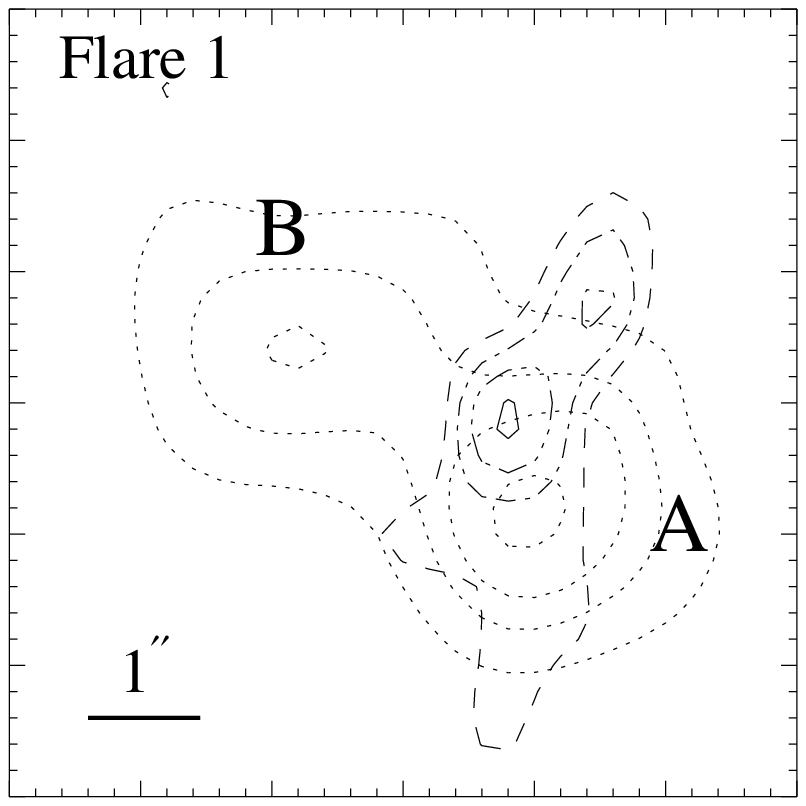}}}
\parbox{4.5cm}{\resizebox{4.5cm}{!}{\includegraphics{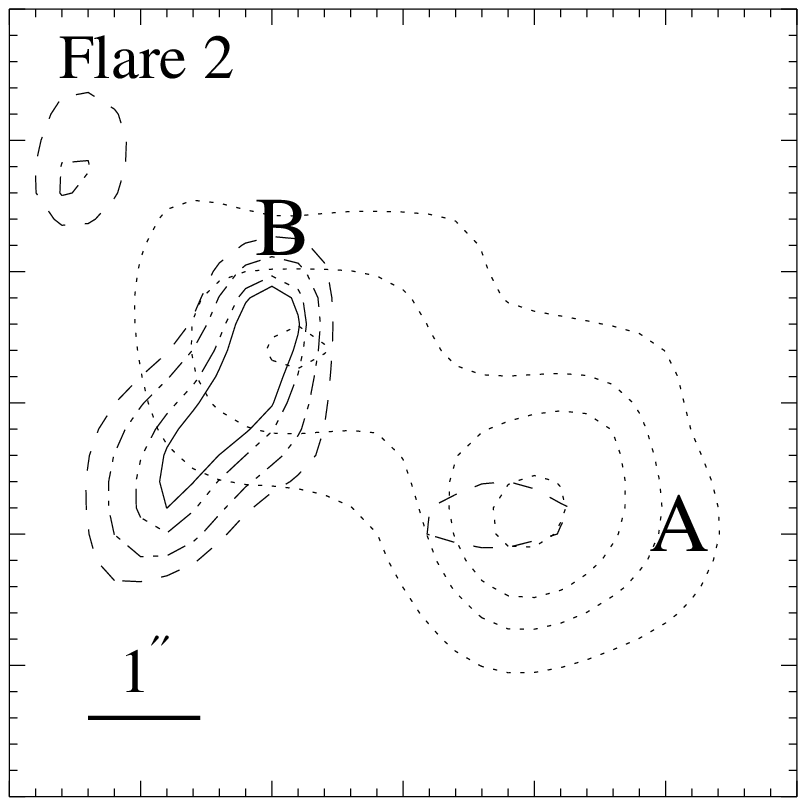}}}
}
\parbox{9cm}{
\parbox{4.5cm}{\resizebox{4.5cm}{!}{\includegraphics{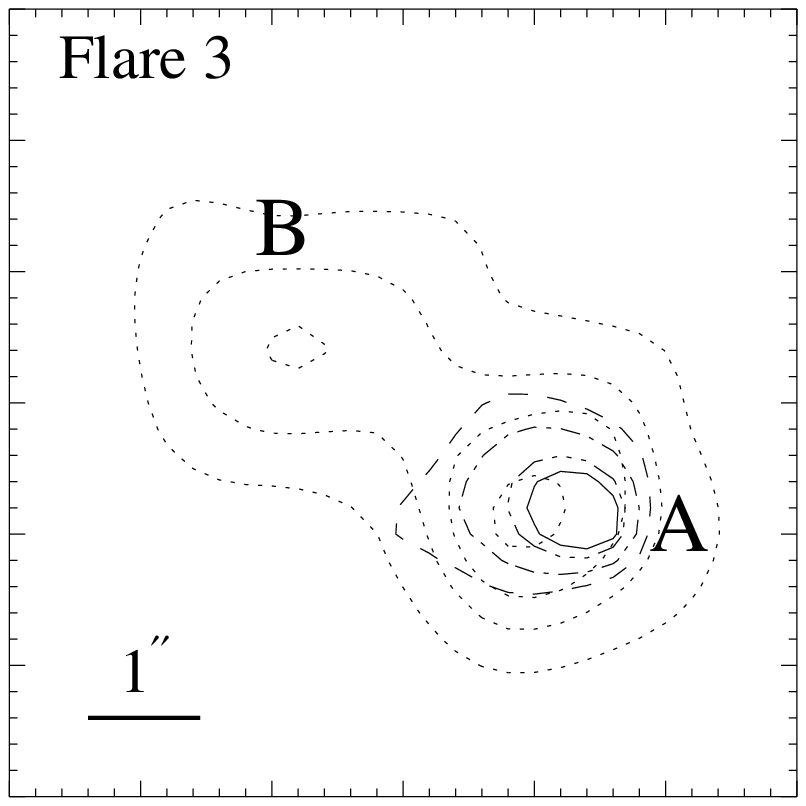}}}
\parbox{4.5cm}{\resizebox{4.5cm}{!}{\includegraphics{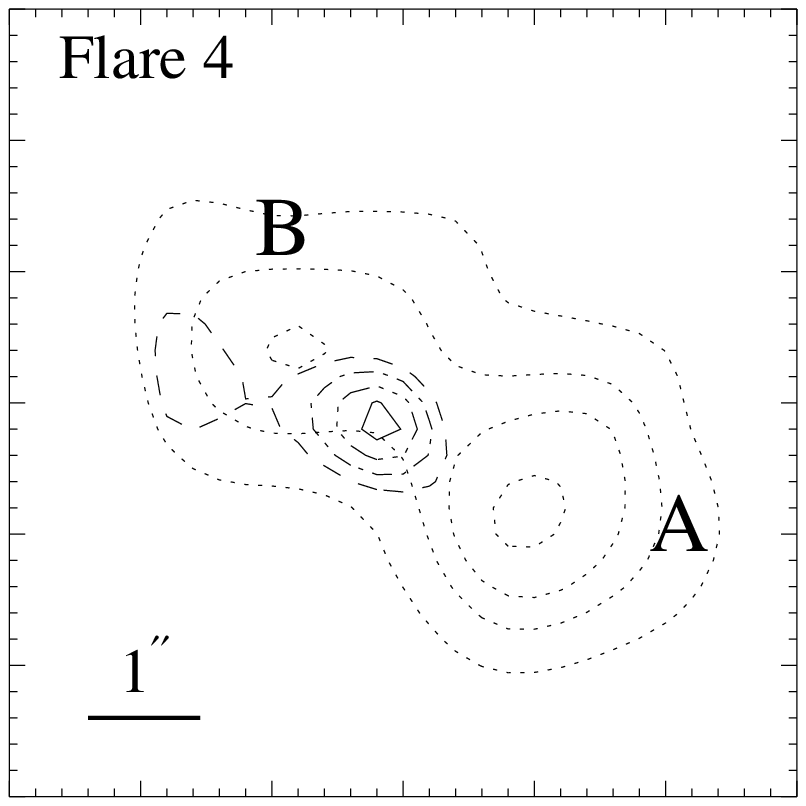}}}
}
\caption{Contour plot of EPIC MOS\,1 images at the position of Castor\,AB during the four flares (see lightcurve Fig.~\ref{fig:lcs}). Underlined dotted contours represent the quiescent emission, and mark the position of Castor\,A and Castor\,B. Contours are normalized to the peak. The relative shift of the photon center indicates that the major part of the emission moved from Castor\,A to Castor\,B and vice versa.}
\label{fig:mos_images}
\end{center}
\end{figure}

The second to fourth flare observed with EPIC can be assigned to either one 
of Castor\,A or~B because of the simultaneous coverage with {\em Chandra} 
(see Fig.~\ref{fig:lcs} and Table~\ref{tab:flareparams}). 
In order to determine the host of the first flare, which was observed by
{\em XMM-Newton} alone, we make use of EPIC MOS images: we extract MOS\,1
images for data subsets corresponding to the time of flares. 
A contour plot of each of these images 
is seen in Fig.~\ref{fig:mos_images}. For clarity we have underlaid the
contours for the quiescent emission. 
Their two peaks are separated by $\sim 3.2^{\prime\prime}$ outlining 
the positions of Castor\,A and~B. 
A shift of the photon peak from one flare to the other is noticeable, 
indicating that the emission
comes from different positions within the Castor\,AB system. For flare\,\#2 
to~\#4 the star that is closer to the peak of the flare contours is the one 
which is also listed in Table~\ref{tab:flareparams}, i.e. the MOS image is 
consistent with the {\em Chandra} lightcurves. The contour of flare\,\#1 
suggests that this event took place on Castor\,A. This way we have identified 
the host of all flares.
However, Fig.~\ref{fig:mos_images} makes clear that this analysis puts us at 
the limit of the capability of EPIC. 
%
%

\begin{figure}
\begin{center}
\resizebox{9cm}{!}{\includegraphics{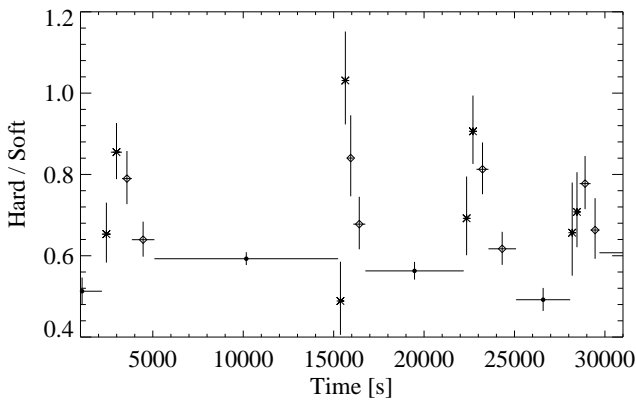}}
\caption{Time evolution of spectral hardness during the four large flares observed with {\em XMM-Newton}: Hard/Soft = counts [0.85-2.0\,keV] / counts [0.3-0.85\,keV]. Different plotting symbols denote different phases of the lightcurve: {\em circles} - quiescence, {\em asterisks} - rise of flare, {\em diamonds} - decay of flare. The bins for the rise phase of all flares are defined from $10-50$\% and $50-100$\% of the peak amplitude, and for the decay phase the other way round. For most flares the maximum hardness, i.e. highest temperature, is reached during the late rise phase of the lightcurve.}
\label{fig:hrs}
\end{center}
\end{figure}

To trace the evolution of the coronal temperature 
during the flares we computed a hardness
ratio derived from the counts in a soft band (S; $0.3-0.85$\,keV) and a hard 
band (H; $0.85-2.0$\,keV) of EPIC pn. 
For this choice of energy bands the Ly$\alpha$ line and the triplet of neon,
which are indicators of hot plasma, are located in the hard band. 
The temporal behavior of $H/S$ is displayed in 
Fig.~\ref{fig:hrs}. 
For three out of the four flares the peak in the spectral hardness 
precedes that of the lightcurve indicating that the outburst is a result of 
heating. Flare\,\#2 is remarkably hard, i.e. the relative 
amplitude of this event in $H/S$ with respect to the other 
flares is larger than in Fig.~\ref{fig:lcs} where count rates in the 
$0.3-5$\,keV broad band are shown. Note that this event is weaker
in the LETGS lightcurve than expected from direct comparison with the EPIC lightcurve. 
It seems natural to explain this difference by a combination of flare hardness, i.e.
temperature, and instrumental sensitivity. 
However, emission at higher energies ($E > 2$\,keV)
where the sensitivity of the LETGS drops sharply is not more prominent than for the 
other flares. 

A time-resolved study of the high-resolution RGS spectrum would provide more information
about temperature and density changes during the flares. However, due to poor statistics 
such an analysis does not seem promising. For Castor\,C which is brighter
in X-rays we found enhanced emission at $\lambda < 16$\,\AA~ (either from continuum or
unresolved lines), and marginal ($1.5\,\sigma$) evidence for variable $G$ and $R$ ratio 
during flares (\cite{Stelzer02.1}). Similarly \citey{Guedel02.1} observed variable
emission line fluxes during a giant flare on Proxima Centauri.    
%
%

\section{X-ray Activity Level}\label{sect:act}

X-ray flux, luminosity, and the ratio between X-ray and bolometric luminosity
are frequently used to measure the strength of stellar activity.  
We computed the X-ray luminosity of 
Castor\,AB by integrating the time-averaged EPIC pn spectrum. 
For the {\em ROSAT} band
we derived $\lg{L_{\rm x}} = 28.96$\,erg/s,
compatible with earlier measurements of {\em EXOSAT} and {\em Einstein} (\cite{Pallavicini90.2}), 
and {\em ROSAT} (\cite{Schmitt94.1}).   

The {\em Chandra} data can 
be used to obtain an independent, estimate of the X-ray luminosity of each
of the two binaries. Since  
the zeroth order does not contain any information on the spectral distribution
of the counts we used PIMMS\footnote{http://asc.harvard.edu/toolkjet/pimms.jsp}  to derive the X-ray luminosities. We find $\lg{L_{\rm x,A}}=28.9$\,erg/s and 
$\lg{L_{\rm x,B}}=28.8$\,erg/s in the {\em ROSAT} band 
assuming a one-temperature Raymond-Smith model of $kT=1$\,keV. 
While clearly too simplistic this rough estimate shows that the result obtained
with both satellites are compatible with each other.   

Based on the {\em Chandra} luminosities the $\lg{(L_{\rm x}/L_{\rm bol})}$ ratio
can be evaluated for all components of Castor\,AB. To compute the bolometric luminosity 
for the A-type primaries we made use of the published $V$ magnitudes (e.g. \cite{Huensch99.1}) 
and the bolometric corrections given by \citey{Kenyon95.1}. 
The bolometric magnitudes of the secondaries were kindly made available by M. G\"udel (priv.comm.).
Assuming that the X-rays are emitted by the primaries leads to 
$\lg{(L_{\rm x}/L_{\rm bol})} = -6.3$ and $-6.0$ for Castor\,A and~B, respectively.
This is higher than the canonical value of $L_{\rm x}/L_{\rm bol} \sim 10^{-7}$
measured for early-type (OB) stars that produce X-rays in their winds. 
If instead the unresolved secondaries are responsible
for the X-ray emission $\lg{(L_{\rm x}/L_{\rm bol})}$ is $-3.7$ and $-3.5$, 
as typical for late-type active stars. Thus the observed emission level suggests the low-mass
companions as X-ray emitters. 

\begin{figure}
\begin{center}
\resizebox{9cm}{!}{\includegraphics{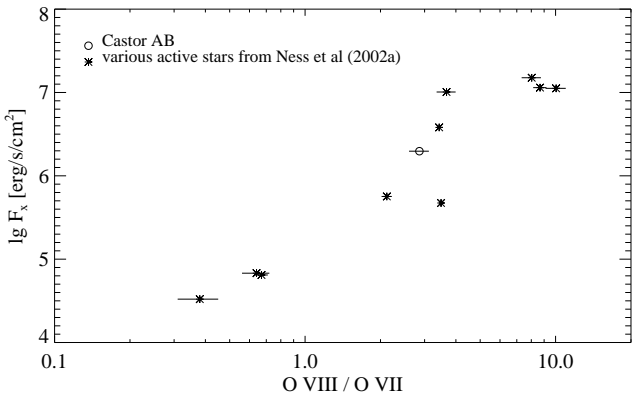}}
\caption{X-ray surface flux versus Ly$\alpha/r$ line ratio of oxygen as a measure for the coronal temperature. $F_{\rm x}$ was computed for the {\em ROSAT} band ($0.1-2.4$\,keV).}
\label{fig:fx_ratio}
\end{center}
\end{figure}

Recently, \citey{Ness02.1} investigated the 
dependence between coronal temperature (expressed in the Ly$\alpha/r$ ratio) 
and activity level (represented by the X-ray surface flux) 
for a number of late-type stars. We add Castor\,AB to this sample. 
When converting $L_{\rm x}$ to flux we assumed (i) that the secondaries are the
source of the X-rays, and (ii) that the observed X-rays are distributed
over the surface of both late-type stars. 
Both assumptions are supported by our discussion above. 
Fig.~\ref{fig:fx_ratio} demonstrates that Castor\,AB are moderately active 
stars that fit well into the straight correlation between X-ray flux and 
temperature characterizing stellar coronae.
%
%

\section{Summary}\label{sect:summary}

Castor is so far the only stellar system observed simultaneously with
both {\em Chandra} and {\em XMM-Newton}. The contemporaneous observation
allowed us to explore the X-ray emission from this multiple star
with unprecedented detail using the complementary capabilities of both
satellites. 

The $\alpha$\,Gem system, composed of the two spectroscopic binaries Castor\,A
and Castor\,B, has been resolved for the first time in X-ray light. At a 
separation of $4^{\prime\prime}$ the only X-ray instrument capable of 
resolving Castor\,A and~B completely is presently {\em Chandra}. 
The {\em Chandra} LETGS allowed us to separate both the 
lightcurves and the high-resolution 
spectrum of the binaries, demonstrating that these stars behave like a typical
late-type coronal X-ray emitter: they exhibit strong variability including 
several flares, and their spectrum is dominated by 
emission lines of O\,VIII, O\,VII, Fe\,XVII, and Ne\,IX. However, the
LETGS spectrum is of low quality (due to the off-axis position and X-ray
faintness of Castor\,A and~B). 

A quantitative analysis of the high-resolution spectrum of Castor\,AB was 
performed with the {\em XMM-Newton} RGS. Its resolution is insufficient
to resolve the two binaries from each other, 
but its higher sensitivity allows to use 
individual emission lines and their flux ratios for coronal diagnostics. 
We measured the temperature sensitive $G$ ratio and the density sensitive
$R$ ratio of the helium-like ions of oxygen and neon, and compared these
ratios to calculations for collisional
ionization equilibrium to learn more about the conditions in the coronal 
plasma of Castor\,AB. 
Simulating a neon-free RGS spectrum based on the temperature structure 
inferred from the EPIC spectrum and subtracting this simulated spectrum from
the data we have shown that the neon triplet is severely
contaminated by a blend of iron lines which affect the $r$ and the $i$ line.
As a result the density derived from the Ne\,IX triplet must be 
regarded as quite uncertain. 

For oxygen we are able to derive more reliable plasma parameters.
For this element the electron temperature is $\sim 2$\,MK and the electron 
density is $(0.5...1)\,10^{10}\,{\rm cm^{-3}}$. The two values for
the density arise from differences of the radiation field of Castor\,A and~B. 
However, the assumption of equal line ratios for both stars may have
introduced an uncertainty of the same order, such that we consider the values
cited above as a likely range for the density rather than two different values
for the coronae of the two stars. 

Another temperature sensitive line ratio compares the strength of the 
Ly$\alpha$ line of an H-like ion to the resonance line of the He-like ion of 
the same element. Representing two
different ionization stages this ratio directly relates to coronal
temperature. We showed that the coronal temperature of Castor\,AB 
as measured by Ly$\alpha/r$ fits well into the correlation with X-ray surface
flux observed for active stars. 
Note that the Ly$\alpha/r$ measurements for the active stars we used for comparison 
(data from \cite{Ness02.1}) were all based
on LETGS data, while our data point for Castor derives from an RGS observation.
Previous studies of stars observed with both the RGS
and the LETGS indicate a tendency of the RGS to measure slightly higher 
values for Ly$\alpha/r$ than the LETGS 
(\cite{Audard01.1}, \cite{Raassen02.1}, \cite{Stelzer02.1}). However, most
of the RGS and LETGS measurements are compatible with each other within the 
statistical uncertainties. Therefore, we consider the respective calibration
of both instruments sufficient for a direct comparison.

The X-ray lightcurves of Castor\,A and Castor\,B show that both are
strongly variable and frequently subject to flaring. With a combination of
{\em Chandra} lightcurves 
and EPIC MOS images for individual flares we established 
the host of each of the six flares: 
{\em Chandra} directly resolves component\,A and~B, while in the MOS
image the photon center shifts during flares. 
Three events occurred on Castor\,A and
three on Castor\,B. Two of the flares were observed only with {\em Chandra}
(during the last part of the observation, when the {\em XMM-Newton} exposure
had already finished). For these two flares the low S/N of {\em Chandra}
does not allow an analysis of its characteristics. For the four flares
that occurred within the observing time of {\em XMM-Newton} we made use of
the excellent sensitivity of EPIC pn, and derived duration and energy output
for each of them. The flare recurrence time can be estimated to one event
every $3-5$\,h. Similarly strong activity was observed on Castor\,AB in an
earlier {\em XMM-Newton} observation presented by \citey{Guedel01.1}.
However, lacking simultaneous high-spatial resolution {\em Chandra} data 
some uncertainty remained as to the origin of the flares (Castor\,A or~B).
On basis of this earlier {\em XMM-Newton} observation \citey{Guedel01.1}
supposed that the flare rate on Castor\,B is higher than on~A, and flares
decay faster than on~A. The observations presented here suggest similar
flare rates for both~A and~B. For the flares observed with {\em XMM-Newton} 
our quantitative analysis shows indeed that decay times are shorter for~B
than for~A. But still better statistics are required to clearly establish a
difference in the flaring properties of Castor\,A and~B.
The flares on all visual components of the Castor system (including YY\,Gem; 
see \cite{Stelzer02.1}) are of short-duration similar to the majority of
X-ray flares seen on young stars, 
but in contrast to the common type of long-duration flares 
on more evolved stars such as RS\,CVn binaries. 
The X-ray luminosity and strength of all X-ray flares observed on 
Castor\,A and~B so far  
is similar to that of dMe flare stars with amplitudes of
a factor of $2...3$ (see e.g. \cite{Pallavicini90.2}). This supports the
common believe that the late-type companions in each of the binaries
are responsible for the X-ray emission.

\begin{acknowledgements}
We are grateful to D. Porquet for help with the model calculations. 
BS wishes to thank M. G\"udel and V. Costa for providing information on 
the stellar parameters and UV flux of Castor\,A and~B, and G. Micela
for careful reading of the manuscript. The CORA software is kindly provided by
J-.U. Ness. 
BS acknowledges financial support from the European Union by the Marie
Curie Fellowship Contract No. HPMD-CT-2000-00013. 
\end{acknowledgements}

\end{document}